\documentclass[structabstract]{aa}  
\usepackage{graphicx}
\usepackage{txfonts}
\usepackage{longtable}
\usepackage{natbib}
\bibpunct{(}{)}{;}{a}{}{,} 

\def\Gb {G16.59$-$0.05}

\def\wat {H$_2$O}
\def\meth {CH$_3$OH}
\def\amm {NH$_3$}
\def\nthp {N$_2$H$^+$}
\def\mcn {CH$_3$CN}
\def\hcop {HCO$^+$}
\def\kms {km s$^{-1}$}
\def\Vlsr {$V_{\rm LSR}$}

\def\mum {$\mu$m}

\def\HII{H{\sc ii}}
\begin{document}
%
   \title{A study on subarcsecond scales of the ammonia and continuum emission 
   toward the \Gb\ high-mass star-forming region}

   \author{L. Moscadelli\inst{1}
          \and
          R. Cesaroni\inst{1}
          \and
          \'A. S\'anchez-Monge\inst{1}
          \and
          C. Goddi\inst{2}
          \and
          R.S. Furuya\inst{3}
          \and
          A. Sanna\inst{4}
          \and 
          M. Pestalozzi\inst{5}
          }

   \institute{INAF-Osservatorio Astrofisico di Arcetri, Largo E. Fermi 5, 50125 Firenze, Italy \\
             \email{mosca@arcetri.astro.it}
          \and   
             Joint Institute for VLBI in Europe, Postbus 2, NL-7990 AA Dwingeloo, the Netherlands 
          \and   
            The University of Tokushima Minami Jousanjima-machi 1-1, Tokushima, Tokushima 770-8502, Japan 
          \and   
             Max-Planck-Institut f\"{u}r Radioastronomie, Auf dem H\"{u}gel 69, 53121 Bonn, Germany                
          \and
           INAF-Istituto Fisica Spazio Interplanetario, Via Fosso del Cavaliere 100, I-00133 Roma, Italy
           }

   \authorrunning{Moscadelli et al.}             
   \titlerunning{A subarcsecond study of \Gb}           
           
   \date{}

 
  \abstract
   {}
   {We wish to investigate the structure, velocity field, and stellar content
    of the \Gb\ high-mass star-forming region, where previous studies have
    established the presence of two almost perpendicular (NE--SW and SE--NW),  massive
outflows, and a rotating disk traced
    by methanol maser emission.}
   {We performed Very Large Array observations of the radio continuum and
   ammonia line emission, complemented by COMICS/Subaru and Hi-GAL/Herschel
   images in the mid- and far-infrared (IR).}
   {Our centimeter continuum maps reveal a collimated radio jet that is oriented E--W
   and centered on the methanol maser disk, placed at the SE border 
   of a compact
   molecular core. The spectral index of the jet is negative, indicating non-thermal
   emission over most of the jet, except the peak close to the maser disk,
   where thermal free-free emission is observed. 
   We find that the ammonia emission presents a bipolar structure consistent
  (on a smaller scale) in direction and velocity with that of the NE--SW
  bipolar outflow detected in previous CO observations. After analyzing 
 our previous \nthp(1--0) observations again, we conclude that two scenarios are possible.
  In one case both the radio jet and the ammonia emission would trace 
  the root of the large-scale CO bipolar outflow.  
  The different orientation of the jet and the ammonia flow could be explained
  by precession and/or a non-isotropic density distribution around the star.
  In the other case, the \nthp(1--0) and ammonia bipolarity is interpreted 
  as two overlapping clumps moving with
  different velocities along the line of sight.    
   The ammonia gas also seems to undergo rotation consistent with
   the maser disk. Our IR images complemented by archival data allow us to
   derive a bolometric luminosity of $\sim$10$^4~L_\odot$  and to conclude that most
   of the luminosity is due to the young stellar object in the hot molecular core.}  
   {The new data suggest a scenario where the luminosity and the outflow activity of the whole region 
   could be dominated by 
   two massive young stellar objects: 1)~a B-type star of $\sim$12~$M_\odot$ at the center of the maser/ammonia
   disk; \ 2) a massive young stellar object (so far undetected), 
   very likely in an earlier stage of evolution than the B-type star,
   which might be embedded inside the compact molecular core and power 
   the massive, SE--NW outflow.} 


   \maketitle
%

\section{Introduction}

The study of high-mass ($M>8~M_\odot$) star formation still faces fundamental
questions.  Among the most important issues still to be clarified are \ 1)~the
role played by accretion disks to convey mass onto the (proto)star; \ 2)~the
properties of the (proto)stellar outflows (neutral vs ionized; wide-angle vs
collimated); \ 3)~the way an \HII~region develops and expands.  From a
theoretical point of view, models of massive star formation are complicated by
the need to consider the effects of the intense stellar radiation of a massive
young stellar object (MYSO), which, by heating, ionizing and exerting pressure
on the circumstellar gas, strongly influences the process of mass accretion and
ejection \citep[e.g.,][]{Pet10,Cun11}. Models predict that an \HII~region is
quenched, remains trapped or expands hydrodynamically, depending on the balance
between the stellar radiation and thermal pressure, on one hand, and the
gravitational pull of the massive (proto)star and the ram pressure of the
infalling material, on the other \citep{Ket02,Ket03}.

 From an observational point of view, the large (typically several kpc) distances
of massive stars and their origin in clusters make it difficult to disentangle
the physical and kinematical properties of a single massive (proto)star from
those of other cluster members.  Prior to the advent of the Atacama Large
Millimeter Array (ALMA), the angular resolution ($\ge$ a few 0\farcs1) of
millimeter interferometers (PdBI, SMA) was in most cases inadequate for resolving
massive accretion disks (with model-predicted sizes of a few hundred AU) and
identify the protostellar outflow emitted from a ``single'' MYSO (hidden in the
complex emission pattern resulting from the interacting outflows of the
cluster).  This can explain the failure to detect disks in O-type stars and the
small number of bona-fide detections obtained for B-type stars
\citep{Ces07,San13b}. Use of the Very Large Array (VLA) has permitted the
detection of weak ($\sim$1--10~mJy), relatively compact (size$\le$1\arcsec),
thermal continuum sources nearby MYSOs, which has been recognized as an
important tool for investigating accretion and ejection phenomena.  The origin of
such emission has been interpreted in terms of photo-ionized spherical
\citep{Pan75} and collimated \citep{Rey86} stellar winds, shock-induced
radiation \citep{Gha98}, and, more recently, trapped \HII~regions \citep{Ket03}.
See also \citet{Rod12} and \citet{San13a}  and references therein, for a list of
different mechanisms proposed to explain the observation of thermal centimeter
continuum emission  in star-forming regions.

With this in mind, in the past years we have focused on the
high-mass star-forming region \Gb, also known as IRAS\,18182--1433.  This object
has been investigated by various authors via thermal line interferometric (SMA,
OVRO, PdBI) and maser VLBI observations \citep[][hereafter B2006, F2008, S2010,
respectively]{Beu06,Fur08,San10a}. Its distance (3.6$\pm$0.3~kpc) has been
recently determined with maser trigonometric parallax observations (Zhang,
private communication). Inside a molecular clump of $\sim$0.5~pc and
$\sim$1900~$M_\odot$ \citep{Beu02d,Hil05}, millimeter continuum and line
observations reveal a number of young stellar objects (YSOs) and a complex flow
pattern, dominated by two almost perpendicular (NE--SW and SE--NW),  massive
outflows (see Fig.~8 of B2006).  18.1~\mum\ emission from the molecular clump 
is marginally detected by \citet{DeB05}
with the 3~m NASA Infrared Telescope Facility.
 The VLA C-array observations at 3.6, 1.3 and
0.7~cm of \citet{Zap06} identify three compact continuum sources, named ``a'',
``b'' and ``c'', within a distance of $\approx$10\arcsec. Sources ``a'' and
``b'' belong to the molecular clump and are separated by $\sim$2\arcsec\ along
the SE--NW direction: source ``a'', the one to NW, is detected only at 0.7~cm,
whereas source ``b'', to SE, is visible both at 3.6 and 1.3~cm.  Interferometric
measurements at 3~mm (F2008) and 1~mm (B2006) show that the continuum emission,
engulfing both ``a'' and ``b'' sources, peaks at the position of source ``a'',
indicating that this source is the most embedded, and plausibly youngest object in
the clump. The SMA detection of high-density molecular tracers (B2006)
with rotational temperatures as high as 130--150~K (inferred from the
CH$_3$CN lines by F2008) close to the position of ``b'', indicate that this
centimeter continuum source is likely associated with a hot molecular core
(HMC).  For the sake of simplicity, hereafter we will refer to source ``a''
as {\it mm-core} and to source ``b'' as {\it HMC}.

Intense ($\ge$10~Jy) Class~II 6.7~GHz and Class~I 44.1~GHz methanol and 22~GHz
water masers have been detected toward the \Gb\ star-forming region. The
European VLBI Network (EVN) observations of S2010 demonstrate that the 6.7~GHz
CH$_3$OH masers are associated with the HMC and trace an
elongated structure of $\sim$2000~AU, with the 3-D maser velocity pattern
suggesting rotation (see Fig.~7 of S2010) about a mass\footnote{The mass quoted
here differs from that computed by S2010 (35~$M_\odot$), 
since we have taken
into account the new distance estimate of $d$=3.6~kpc (instead of 4.4~kpc),
 and because of slightly different assumptions concerning the systemic
velocity adopted and the maser spots used for the calculation 
(2-3 spots might not be tracing the disk).
Our new estimate of \ $\sim$12~M$_{\odot}$ uses a rotation velocity of 
\ 5.1~\kms (for a systemic velocity of 60~\kms) at a radius of 0\farcs11.
} 
of $\sim$12~M$_{\odot}$, assuming centrifugal equilibrium. \citet{Mos11b} find that
most of the 6.7~GHz maser features present a regular internal \Vlsr\ gradient,
which can be also interpreted in terms of Keplerian rotation around a star
having a similar position and mass as derived from the maser 3-D 
velocity distribution. Water masers, monitored with the Very Long Baseline Array
(VLBA) by S2010, appear distributed close to the HMC, with measured proper
motions tracing fast ($\approx$50~\kms) and poorly collimated expansion to the
west (see Fig.~5b of S2010). A sketch of the main
emission features in the \Gb\ star-forming region is presented in Fig.~8 of S2010, 
which illustrates the spatial
distribution of the continuum sources, and thermal and maser line tracers.

This paper reports on new observations of \Gb\ performed with the VLA and Subaru
telescope at radio and infrared wavelengths. We have imaged both the continuum
and ammonia line emission at subarcsecond resolution and complemented these
data with Herschel continuum images from the Hi-GAL survey. Our purpose is to
investigate the structure of the molecular and ionized gas close to the origin
of the large scale molecular outflow(s) and establish the number and nature of
the MYSOs associated with the maser disk and/or outflow(s). In the following,
after illustrating the observations in Sect.~\ref{obs}, 
our findings will be presented and discussed in Sects.~\ref{sjet},~\ref{lum}~and~\ref{ammo}
and a scenario will be eventually proposed in Sect.~\ref{sscen} to explain in a consistent
manner all the features of this intriguing high-mass star-forming region.
 Finally, conclusions are drawn in Sect.~\ref{conclu}.

\section{Observations}
\label{obs}

\subsection{Very Large Array}

\subsubsection{\amm\ emission and \wat\ masers}
\label{obs-nh3}

We observed the \amm(1,1) (at rest frequency of 23694.5060~MHz) and \amm(2,2)
(at 23722.6336~MHz) main lines, and the water maser emission (at 22235.08~MHz)
in \Gb\ with the NRAO VLA\footnote{The National Radio Astronomy Observatory is operated by Associated Universities,
Inc., under cooperative agreement with the National Science Foundation.}
in the B--Array configuration (project code 12A-054) in three different runs of two hours
on July 5, July 30, and August 6 2012. We recorded dual polarization using two IFs
(each comprising eight adjacent 4~MHz bandwidths),  one centered at a rest frequency
(23708.57~MHz) halfway between the \amm (1,1)~and~(2,2) main lines, and the
other at the water maser sky frequency. Considering the VLA correlator
capabilities in Summer 2012, this frequency setup was selected to observe both
the \amm\ and the water maser emission, to maximize the number of observable
\amm (1,1)~and~(2,2) satellite lines and, at the same time, attain a high enough
velocity resolution to resolve the \amm\ lines. We could observe the \amm (1,1)
satellite lines with velocity separation (from the main line) of $\pm$7~\kms\
and +19~\kms, and the \amm (2,2) satellite lines at $+$16~\kms\ and $+26$~\kms.
Correlating each 4~MHz bandwidth with 128 channels, we achieved a velocity
resolution of 0.39~\kms.

The primary and phase calibrators were the VLA calibrators 3C286 and J1832-1035,
respectively.  The phase calibrator is separated from \Gb\ on the sky by \
4\fdg8. Each 2-hour run included a single 5-min scan on the primary calibrator,
eight 3-min scans on the phase calibrator, and a total of 1.1 hours of on-source
time. First, we applied the amplitude and phase corrections derived from
the calibrators to the strongest (reference) water maser channel, and then, we
self-calibrated the reference maser channel. Self-calibration was effective in
improving the signal-to-noise ratio (SNR) of the image of the reference maser
channel by a factor \ $\ge$10. For each of the three observing runs, the
calibration of the \amm\ line, water maser and continuum emission of \Gb\
consisted of two steps, applying first the calibrator amplitude and phase
corrections, and then the self-calibration solutions from the reference maser
channel. The line data have been also corrected for the time variable Doppler
shift (not accounted by the VLA correlator), which, during a single run, causes
the line emission to drift by at most the velocity
resolution channel.

Before imaging, the calibrated data of the three runs have been concatenated. We
produced naturally-weighted and $uv$-tapered (Gaussian tapering with FWHM of
400~k$\lambda$) images  of the \amm\  main and satellite lines, water maser and
continuum emission. The continuum data was obtained by averaging the 13 (out of
the 16) 4~MHz bandwidths of the two IFs not containing \amm (1,1)~and~(2,2), nor
the water maser lines. 
The naturally-weighted beam has a FWHM size of \ 0\farcs48$\times$0\farcs31, at
PA = 7$\degr$, with small difference between the \amm\ line and continuum maps.
The continuum emission map, restored using a round beam with FWHM size of \
0\farcs4, has a rms noise level of \ 0.06~mJy~beam$^{-1}$. To increase the SNR
on the \amm\ lines, before imaging, the data have been smoothed in velocity,
degrading the velocity resolution to 0.8~\kms. The typical rms noise on the
channel maps of the \amm\ main and satellite lines is of \
1.3--1.5~mJy~beam$^{-1}$.

\subsubsection{Continuum}
\label{obs-cont}

We used the VLA A--Array (code: 12B-044) to observe the continuum emission of
\Gb\ at C-, Ku- and K-band  (centered at 6.2, 13.1 and 21.7~GHz, respectively)
in four different runs on October 2, October 21 and
December 30 2012, and January 17 2013. \Gb\ was observed for 15~min at C- and
Ku-band, and for 30~min at K-band. At all bands, the primary and phase
calibrators were the VLA calibrators 3C286 and J1832-1035, respectively.  We
employed the new capabilities of the WIDAR correlator which permits us to record
dual polarization  using four IFs each comprising eight adjacent 128~MHz subbands,
achieving a total bandwidth per polarization of 2~GHz. At 6~GHz and 22~GHz, we
centered one IF at the frequency of the methanol and water maser, respectively.
Correlating each 128~MHz subband with 128 channels, we achieved a velocity
resolution of 45~\kms\ and 13~\kms\ at 6~GHz and 22~GHz, respectively. Since the
methanol and the water maser signals in \Gb\ are  strong
($\approx$20~Jy~beam$^{-1}$) and relatively wide ($\approx$10~\kms), we could
detect them with high ($\ge$100)  SNR even with these coarse velocity
resolutions.

At Ku-band, the amplitude and phase corrections of the \Gb\ visibilities have
been derived working only with the calibrators. At C- and K-band, we first
applied the calibrator corrections to the maser channel and self-calibrated the
maser emission (improving the SNR of the maser image by a large factor
($\ge$10)), and then, before mapping, applied the self-calibration solutions
from the maser to the continuum data.
In each band, the rms noise level of the continuum image is close to the
expected thermal noise: 8/9~$\mu$Jy~beam$^{-1}$ at 6/14~GHz and
23~$\mu$Jy~beam$^{-1}$ at 22~GHz.  The FWHM size of the naturally-weighted beams
at the different observing wavelenghts are listed in Table~\ref{tabfl}.

\subsection{Subaru}

Using the mid-infrared imaging spectrometer \citep[COMICS;][]{Kat00}
at the Cassegrain focus of the 8.2~m Subaru Telescope, we carried
out imaging observations of the 24.5~\mum\ emission toward \Gb\ on
2008 July 15. For this purpose, we configured the Q24.5 filter, and
employed chopping-and-nodding mode for subtracting the
sky-background emission. The camera provides a field of view of
$\sim 42\arcsec\times 32\arcsec$ with a pixel size of 0\farcs13.
Flux calibration was performed toward four standard sources listed
in \citet{Coh99}: HD146051, HD186791, HD198542, and HD198542. 
We estimated the overall uncertainty in the flux
calibrations to be less than 10\%.

To perform the astrometric calibration of the image, we have compared the
Subaru image with the MIPSGAL image at 24~$\mu$m \citep{Car09},
after smoothing the former to the same angular resolution (6\arcsec) as the
latter. We have then aligned the two images by visually overlaying the non-saturated
part of the MIPSGAL image on the corresponding part of the Subaru image.
 The original Subaru image was corrected by
  about 3\arcsec.  
This process, albeit not very accurate, should result in an absolute positional
uncertainty $\la$1\arcsec, comparable to the astrometric accuracy of MIPSGAL.

\section{Continuum emission: A bipolar jet}
\label{sjet}

\begin{figure}
\centering
\resizebox{8.5cm}{!}{\includegraphics[angle=-90]{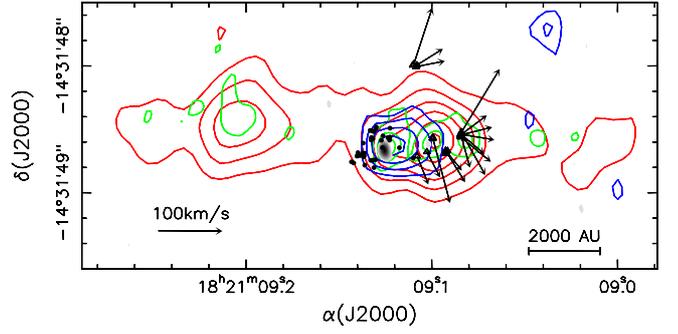}}
\caption{
 Overlay of the radio continuum emission toward the HMC in \Gb\ at 6~cm (red contours),
 2~cm (green), 1.3~cm B-array (blue), and 1.3~cm A-array (gray scale). The
 contour levels range, respectively, from 0.0264 to 0.1584 in steps
 of 0.0264~mJy~beam$^{-1}$, from 0.03 to 0.09 in steps of 0.03~mJy~beam$^{-1}$, from
 0.174 to 0.348 in steps of 0.058~mJy~beam$^{-1}$, and from 0.075 to 0.25 in steps
 of 0.025~mJy~beam$^{-1}$. The solid circles and empty triangles represent,
 respectively, the \meth\ and \wat\ masers from S2010, while
 the arrows are the proper motions of the \wat\ masers.
 The scale for the proper motion amplitude 
 is given on the bottom left of the panel.
 The size of the VLA beams for the different array configurations and wavelengths 
 are reported in Table~\ref{tabfl}.
}
\label{fjet}
\end{figure}

 From previous studies we know that radio continuum emission has been detected
in several locations in the \Gb\ region, but in the present study we will
focus only
on the emission associated with the HMC.
Figure~\ref{fjet} presents a comparison between the continuum maps obtained
by us at three different wavelengths and with different resolutions.

\begin{figure}
\centering
\includegraphics[width=8.5cm]{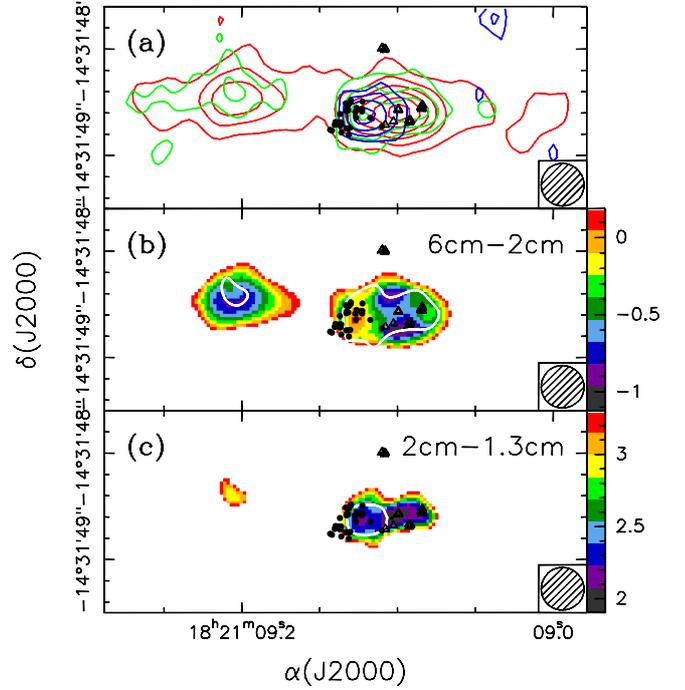}
\caption{
 {\bf a.} Overlay of the VLA maps of the radio continuum emission at 6, 2,
 and 1.3~cm obtained using the common uv range and the same clean beam (shown
 in the bottom right) at all wavelengths. Contours and symbols have the same
 meaning as in Fig.~\ref{fjet}.  The contour levels range from
 0.0264 to 0.1584 in steps of 0.0264~mJy~beam$^{-1}$ at 6~cm, from 0.03 to 0.12 in
 steps of 0.03~mJy~beam$^{-1}$ at 2~cm, and from 0.174 to 0.348 in steps of
 0.058~mJy~beam$^{-1}$ at 1.3~cm.
 {\bf b.} Map of the spectral index between 6 and 2~cm, computed where
 the 6~cm emission is $>5\sigma$. The white contour encompasses the region
 where also the 2~cm emission is $>5\sigma$. The values of the spectral index
 lying outside the white contour are upper limits.
 {\bf c.} Same as panel b, for the spectral index between 2 and 1.3~cm.
}
\label{fjsi}
\end{figure}

For the sake of comparison, we also show the \wat\ and \meth\ maser spots. The
striking result is that, while the 1.3~cm emission appears compact and almost
coincident with the \meth\ masers, consistent with \citet{Zap06}  and S2010, at
longer wavelengths (2 and 6~cm) the emission appears elongated E--W on both
sides of the \meth\ maser spots, extending over $\sim$4\arcsec\ (0.07~pc). This
morphology is strongly suggestive of a bipolar jet, whose powering source could
lie close to the location of the \meth\ masers, near the geometrical
center of the jet. In Table~\ref{tabfl}, we report the peak and integrated
fluxes, derived from naturally-weighted maps, for the VLA A and B array
observations. The thermal and non-thermal fluxes refer to the compact and
extended continuum components, respectively, and are obtained by integrating the
continuum emission inside and outside, respectively, the area where the 1.3~cm
emission is $>5\sigma$ (delimited by the white contour of Fig.~\ref{fjsi}c). The
outer limit of the area where the non-thermal component has been computed, is
determined by the $5\sigma$ level of the 6~cm emission (see Fig.~\ref{fjsi}b).

\begin{table*}[ht!]
\caption{\label{t:fluxes}Centimeter continuum fluxes for the VLA A and B-Array observations.}
\centering
\label{tabfl}
\begin{tabular}{c c c c c c c c c c}
\hline\hline

&$\theta_\mathrm{beam}$
&P.A.$_\mathrm{beam}$
&rms
&$I_\mathrm{peak}$
&$S_\nu^\mathrm{thermal}$
&$S_\nu^\mathrm{non-thermal}$
\\
Wavelength
&(\arcsec$\times$\arcsec)
&(\degr)
&(mJy~beam$^{-1}$)
&(mJy~beam$^{-1}$)
&(mJy)
&(mJy)
\\
\hline
6.0~cm---A		&$0.540\times0.330$	&$-$6	&0.008	&0.164	&$0.031\pm0.002$	&$0.434\pm0.015$	\\
2.0~cm---A		&$0.241\times0.163$	&$-$14	&0.009	&0.117	&$0.099\pm0.005$	&$0.538\pm0.018$	\\
1.3~cm---B		&$0.400\times0.400$	&$+$0	&0.060	&0.375	&$0.146\pm0.008$	&$\le 0.612$		\\
1.3~cm---A		&$0.131\times0.075$	&$-$173	&0.023	&0.265	&$0.324\pm0.032$	&$\le 0.631$		\\
\hline
\end{tabular}
\end{table*}

Besides the morphology, a typical signature of thermal jets is their spectral
index, ranging from $-$0.2 to $+$1.5 \citep[see][]{Ang96} in the radio regime.
To calculate the spectral index all over the continuum emission in \Gb,
we have created maps of the 6, 2, and 1.3~cm continuum
emission using the same $uv$ range for all bands, as well as the same clean
beam (0\farcs4 $\times$ 0\farcs4). In this way it is possible to compare the emission at different
wavelengths minimizing the effects of different samplings of the uv plane and
thus obtaining a more reliable estimate of the spectral index.  The new maps
are shown in the top panel of Fig.~\ref{fjsi}, while in the other two panels
we show the maps of the spectral index $\alpha$ (assuming
$S_\nu\propto\nu^\alpha$) between 6 and 2~cm (Fig.~\ref{fjsi}b) and between 2
and 1.3~cm (Fig.~\ref{fjsi}c). For these two estimates the uncertainty is,
respectively, $\pm$0.2 and $\pm$0.5.  Note that $\alpha$ has been calculated
only where the 6~cm (for Fig.~\ref{fjsi}b) and 2~cm (for Fig.~\ref{fjsi}c)
fluxes are $>5\sigma$. Moreover, the white contour encompasses the region
where the 2~cm (for Fig.~\ref{fjsi}b) and 1.3~cm (for Fig.~\ref{fjsi}c)
fluxes are $>5\sigma$. This means that the values of the spectral indices
lying outside the white contour are upper limits.

The striking result is that the spectral index between 6 and 2~cm is $<$~$-$0.5
over a significant fraction of the region where 6~cm emission is detected. 
 Such negative values
are inconsistent with free-free emission and strongly suggestive of non-thermal
emission. Indeed, this is not the first example of a radio jet
with negative spectral index,
associated with a massive (proto)star.
Synchrotron emission has been proposed to explain the radio emission
in W3(\wat) \citep{Rei95} and, more recently, in HH~80--81
\citep{Car10}, indicating that magnetic fields
may play an important role in shaping such jets and, more in general,
the accretion/ejection process in massive (proto)stars.
In the case of \Gb\ we find that: (i)~the value of
$\alpha$ in Fig.~\ref{fjsi}b is negative all over the region, but becomes
$>-0.1$ (i.e., consistent with free-free emission) close to the \meth\ masers;
(ii)~the spectral index is $\sim$2 between 2 and 1.3~cm (see Fig.~\ref{fjsi}c),
as expected for optically thick free-free emission. 
However, at larger distance from the \meth\ masers (outside the white contour of
Fig.~\ref{fjsi}c), the value of spectral index between 2 and 1.3~cm is
only an upper limit and could be as low as measured between 6 and 2~cm.
Closer to the \meth\ masers, given the low brightness
temperature measured at 1.3~cm (69$\pm$6~K with the A-array), it is unlikely
that the emission is optically thick and we believe that values of $\alpha$
as large as $\sim$2 may be partly due to the uncertainty caused by the
proximity between the two wavelengths used for the calculation.

Nonetheless, it is clear that the difference between $\alpha$(6--2~cm) and
$\alpha$(2--1.3~cm) is real. A possible explanation is that the radio emission
is made of two contributions, with synchrotron dominating at
long wavelengths and at larger separations from the powering MYSO,
and bremsstrahlung preponderant close to the MYSO.
Note infact that, near the
\meth\ masers, $\alpha$ is consistent with free-free at {\it all}
frequencies. Interestingly, this is also where the most compact 1.3~cm
emission is detected with the highest angular resolution (see gray scale in
Fig.~\ref{fjet}), which suggests that this compact emission could be tracing
the dense ionized gas in the neighborhoods of the star powering the jet.

What is the origin of the free-free emission? This could certainly arise from
material ionized by shocks along the jet. However, one cannot exclude that
ionization is due to the stellar photons. In this case, assuming optically
thin emission, from the 1.3~cm continum flux measured with the B-array
(0.69~mJy) one obtains a stellar Lyman continuum of $\sim$10$^{45}$~s$^{-1}$,
corresponding to a zero-age main-sequence (ZAMS) luminosity of
$\sim$4000~$L_\odot$ and a stellar mass of $\sim$10~$M_\odot$
\citep[see][]{Dav11}. According to \citet[][see their Fig.~5]{Beu02b}, 
this implies a mass loss rate for the associated outflow of a few
times $10^{-3}~M_\odot$yr$^{-1}$. It is questionable whether this value can
be compared to those of the two CO outflows imaged on a larger scale by
B2006, F2008, and \citet{Lop09}, because the direction
of the radio jet differs by $\sim$45\degr\ from that of either outflow. It is
hence possible that three massive objects are powering three distinct
jets/outflows in the \Gb\ region. We will come back to this issue in the
following.

While the structure of this star-forming region is clearly very complex
and will be further discussed in the next sections, we may conclude that
our new radio continuum images strongly support the interpretation of the
\meth\ masers as tracing a rotating disk (see S2010), because the radio jet
appears to originate from a MYSO located close to the center of the maser
spots and
almost perpendicular
to the elongated maser cluster.

\section{Bolometric luminosity and its origin}
\label{lum}

\begin{figure}
\centering
\resizebox{8.5cm}{!}{\includegraphics[angle=-90]{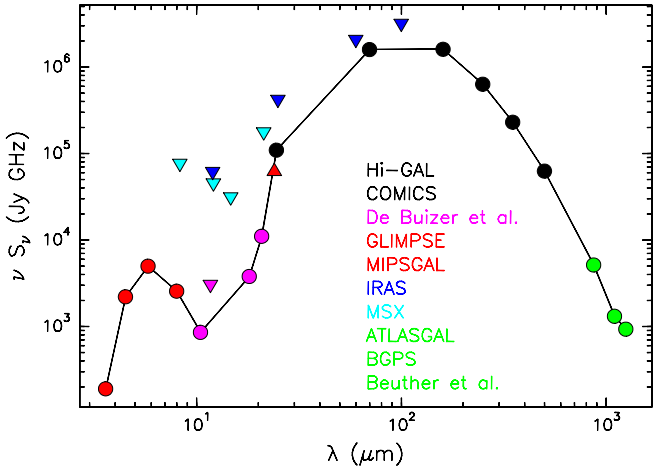}}
\caption{
 Spectral energy distribution of \Gb\ obtained from our IR measurements
 and archival data (see text).
 Triangles denote upper and lower limits.
 The luminosity estimate ($\sim$ $10^4~L_\odot$) has been obtained by integrating
 the SED under the solid line.
}
\label{fsed}
\end{figure}

Given the presence of multiple objects in \Gb, it is important to establish
which of these are the main contributors to the luminosity of this 
star-forming region. We have reconstructed the spectral energy distribution (SED)
using the following archival data: Galactic Legacy Infrared Mid-Plane Survey Extraordinaire \citep[GLIMPSE;][]{Ben03}, 
MIPSGAL/Spitzer 24~$\mu$m \citep{Car09}, 
MSX \citep{Ega03},
IRAS \citep{Neu84}, 
the APEX Telescope Large Area Survey of the Galaxy \citep[ATLASGAL\footnote{The ATLASGAL project is a collaboration between the Max-Planck-Gesellschaft, the European Southern Observatory (ESO) and the Universidad de Chile.};][]{Sch09,Con13}, 
Bolocam Galactic Plane Survey \citep[BGPS;][]{Dro08},
data from the literature \citep{Beu02d,DeB05},
and recent PACS and SPIRE continuum images
obtained in the context of the Herschel
infrared Galactic Plane Survey \citep[Hi-GAL;][]{Mol10b, Mol10a}. 

The flux density estimates of IRAS, MSX, GLIMPSE, BGPS, and ATLASGAL
were taken from
the corresponding source catalogues, while those of Hi-GAL and MIPSGAL
were computed from the images with aperture photometry, by integrating the emission inside
suitable polygons encompassing our source and subtracting the background. Note that the MSX and
IRAS flux densities are considered upper limits, because the
HPBWs encompass a region significantly larger than that of interest for us.
The MIPSGAL flux is a lower limit, because the image is partly saturated.
We stress that such a limit is consistent with our Subaru measurement.

To estimate the bolometric luminosity,
the resulting SED (shown in Fig.~\ref{fsed}) has been
integrated by interpolating linearly between the known fluxes
(solid line in the figure). We obtain $\sim$ $10^4~L_\odot$,
corresponding to a single star of
$\sim$13~$M_\odot$ \citep[see][]{Dav11}.

\begin{figure}
\centering
\includegraphics[width=8.5cm]{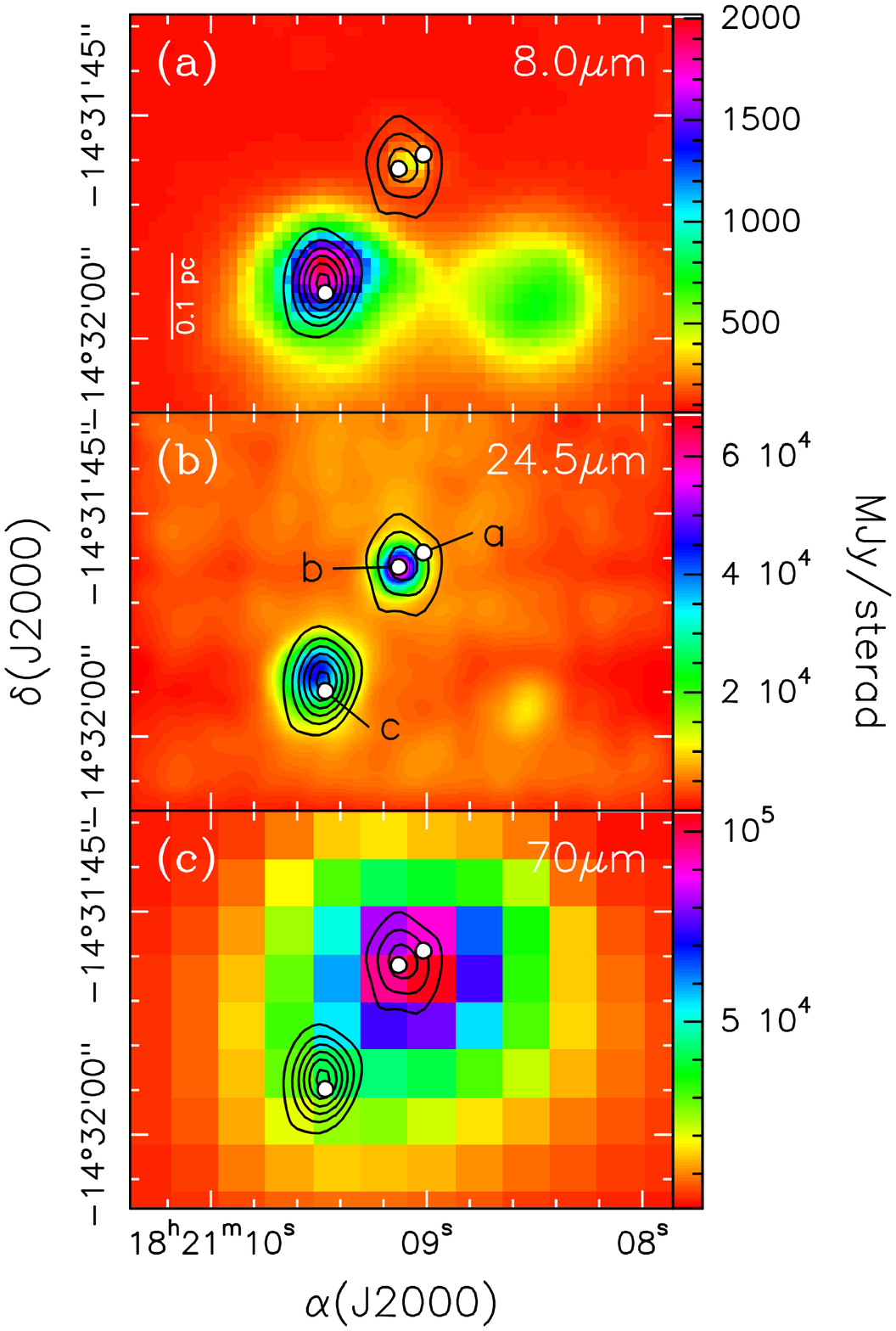}
\caption{
 {\bf a.} Map of the continuum emission at 3.6~cm (black contours; from
 S2010) overlaid on an 8.0~$\mu$m image (resolution 2\arcsec) of the
 \Gb\ region from the GLIMPSE/Spitzer survey \citep{Ben03}.
 {\bf b.} Same as panel a, for our Subaru 24.5~$\mu$m image smoothed to the
 same angular resolution as the 8.0~$\mu$m image (2\arcsec).
 {\bf c.} Same as panel a, for the Hi-GAL/Herschel \citep{Mol10a} 70~$\mu$m image
 (resolution $\sim$10\arcsec).
 The white dots indicate the positions of the centimeter sources ``a''
 (alias mm-core),
 ``b''
 (alias HMC)
 and ``c''
 detected by \citet{Zap06}.
}
\label{firim}
\end{figure}

\begin{figure}
\centering
\resizebox{8.5cm}{!}{\includegraphics[angle=0]{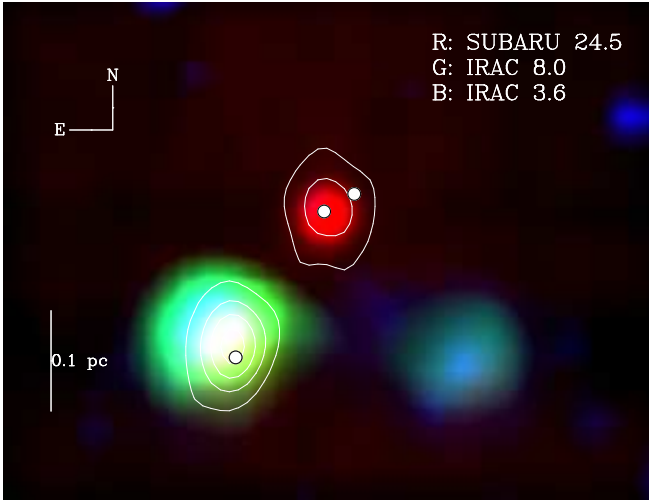}}
\caption{
 Composite color image of the same field shown in Fig.~\ref{firim}, where
 blue is associated with the GLIMPSE/Spitzer 3.6~$\mu$m image, green with
 the GLIMPSE/Spitzer 8.0~$\mu$m image, and red with our 24.5~$\mu$m Subaru
 image. The reddest and hence most extincted source is the one to the north,
 coinciding with the HMC. The VLA 3.6~cm continuum (from S2010) is shown with
 white contours. The white dots have the same meaning
 as in Fig.~\ref{firim}.
}
\label{fcol}
\end{figure}

In the case of \Gb, the single-star assumption is inadequate to describe the
situation, as three prominent IR sources are clearly detected over less than
$\sim$20\arcsec\ (i.e., $\sim$0.3~pc). This is demonstrated by the 8~$\mu$m
GLIMPSE/Spitzer image, as well as our 24.5~$\mu$m Subaru image, shown in
Figs.~\ref{firim}a and~\ref{firim}b, respectively. The problem is that the angular
resolution at $\lambda>70~\mu$m -- i.e., close to the peak of the SED (see
Fig.~\ref{fsed}) -- is not sufficient to resolve the three sources and find
out which is responsible for most of the bolometric luminosity. To
investigate this issue, in Fig.~\ref{fcol} we present a composite color
image of the same region displayed in Fig.~\ref{firim}, where the blue,
green, and red components are associated, respectively, with the 3.6, 8, and
24.5~$\mu$m images (the first two from GLIMPSE/Spitzer).  There is little
doubt that the reddest and hence most embedded object is the northern source,
namely the one coinciding with the HMC. Visual comparison of the three panels
in Fig.~\ref{firim} leads to the same conclusion, as the northern source
becomes progressively more prominent with increasing wavelength, until it
dominates the whole IR emission at 70~$\mu$m.

We conclude that in all
likelihood the HMC is the source responsible for most of the
$\sim$10$^4~L_\odot$ measured over the \Gb\ region.
However, one should keep in mind that, although the HMC is contributing to
most of the IR emission (and hence of the bolometric luminosity), younger
MYSOs might be embedded in the cold, dense surroundings of the HMC, possibly
contributing significantly to the outflow activity in the region. We will come
back to this point later.

\section{Ammonia emission}
\label{ammo}

\begin{figure}
\centering
\resizebox{8.5cm}{!}{\includegraphics[angle=0]{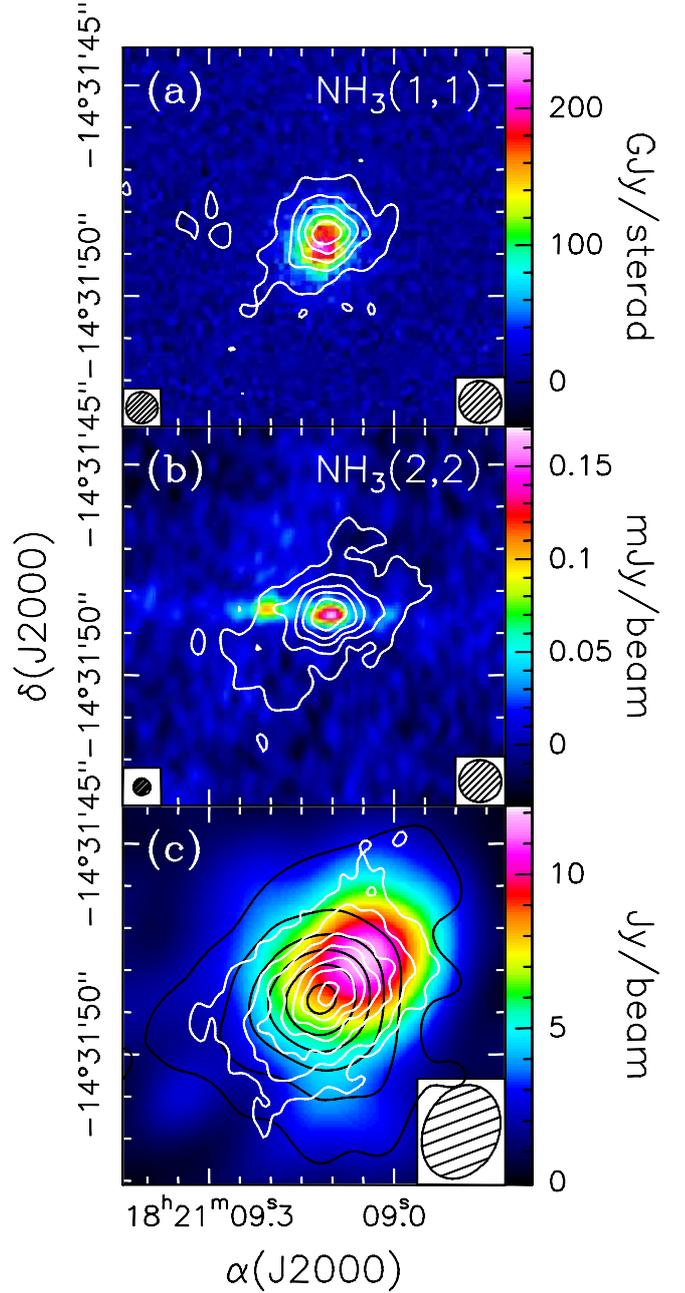}}
\caption{
 {\bf a.} Map of the \amm(1,1) emission averaged over the main line (white
 contours) overlaid on an image of the 24.5~$\mu$m continuum emission obtained
 with Subaru. The circles in the bottom left and right indicate, respectively,
 the angular resolution of the IR and radio data.
 {\bf b.} Same as panel a, for the \amm(2,2) line (contours) and the 6~cm radio
 continuum map (background image).
 {\bf c.} Overlay between the \amm(2,2) (white contours), \mcn(5--4) (black
 contours), and 3.3~mm continuum (image) maps. The ammonia map has been
 smoothed to the same angular resolution (ellipse in the bottom right) as the
 methyl cyanide data (from F2008).
}
\label{famm}
\end{figure}

Our ammonia observations provide us with information that is
helpful to shed light on the structure and velocity field in this region.
Such information is summarized by Fig.~\ref{famm}, which shows three overlays
between the \amm\ line emission and other relevant tracers.

In particular, in Fig.~\ref{famm}a the integrated \amm(1,1) line map is
compared to the 24.5~$\mu$m image obtained with Subaru, which confirms that
the IR emission is tightly associated with the HMC, within the positional
uncertainty. It is also interesting to compare the 6~cm emission tracing the
radio jet with the ammonia emission, as done in Fig.~\ref{famm}b.
The strongest 6~cm continuum peak, which likely traces shocked gas
of the radiojet (see Fig.~\ref{fjet} and Sect.~\ref{sjet}), appears to coincide with the \amm\ peak, while both
are offset by $\sim$0\farcs5 ($\sim$1800~AU) to the west of the 1.3~cm continuum
peak and \meth\ masers -- which likely pinpoint the position of the star powering the radio
jet (see Fig.~\ref{fjet}). This fact strongly suggests that
the jet is impinging against dense molecular gas to the west, where the
brightest radio emission is seen. This hypothesis is supported by the
distribution and proper motions of the \wat\ masers (see Fig.~\ref{fjet}),
which are known to be associated with shocks. The proposed scenario could
also explain why the jet appears slightly more extended to the east, as in
this direction the gas would be much less dense.

Figure~\ref{famm}c sheds further light on the structure of the region.
This
presents an overlay between our map of the
\amm(2,2) inversion line and the maps of the \mcn(5--4) line and 3.3~mm
continuum emission from F2008. Clearly, ammonia and methyl cyanide trace
the same region (the HMC), as expected, while both are significantly offset
(by $\sim$1\farcs3 or $\sim$0.022~pc) from the peak of the mm continuum,
i.e., from the mm-core,
as pointed out by B2006 and F2008.

All these features can be explained by a scenario where a massive star has
formed close to the border of a dense molecular core (the mm-core) and has
heated the surrounding gas, thus creating a HMC. This explains the offset
between the HMC and the mm-core. The latter must be quite massive, because it
appears to dominate the mm continuum emission despite its low temperature
($\sim$38~K; see \citet{Wil05}) with respect to the HMC
($\ga$130~K; see F2008). Using the same dust properties and continuum flux
density (37~mJy) as in F2008, one can estimate a mass of
$\sim$200~$M_\odot$ for the mm-core. Such a large value suggests that
other MYSOs, possibly in a much earlier evolutionary phase than the star in
the HMC, might be forming inside the mm-core.

\subsection{Morphology and kinematics of the \amm\ emission}
\label{sammo}

\begin{figure}
\centering
\resizebox{8.5cm}{!}{\includegraphics[angle=0]{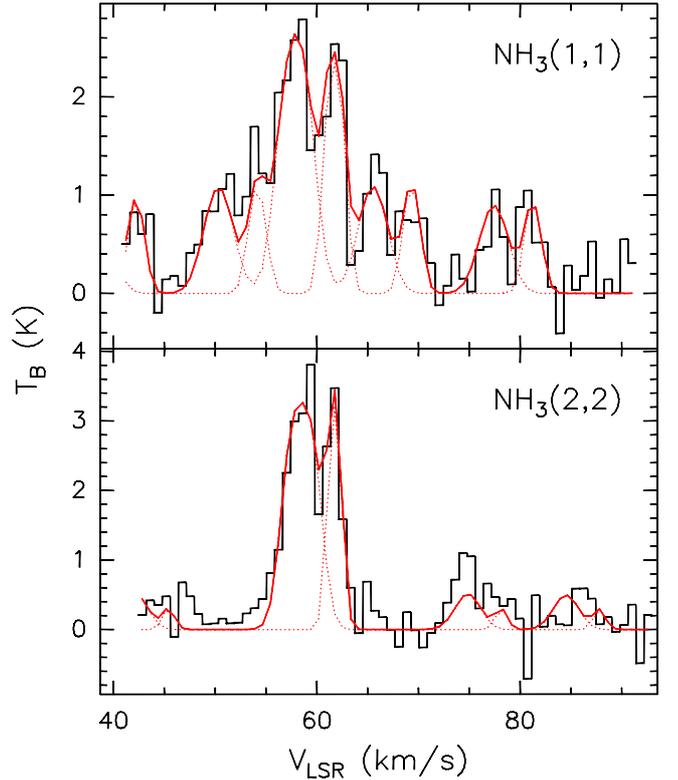}}
\caption{
 Spectra of the \amm(1,1) (top) and (2,2) (bottom) inversion transitions
 obtained by averaging the emission over the HMC. The red solid curves are
 multiple Gaussian fits taking into account the hyperfine structure of
 the lines and two distinct velocity components (at about \ 58.2~\kms\ and 61.8~\kms). 
 The red dotted curves
 indicate the contribution of each velocity component to the total fit.
}
\label{fspamm}
\end{figure}

In Fig.~\ref{fspamm}, we show the average spectra of the \amm(1,1) and (2,2)
inversion transitions over the HMC. The most intriguing feature is the
presence of a double peak, clearly detected in both transitions. The fact
that such a double peak is seen also in the optically thin satellites proves
that this is not the effect of self-absorption, but is due to two distinct
velocity components. The red profiles in Fig.~\ref{fspamm} are fits to the
\amm\ lines taking into account the hyperfine structure of the transitions,
and clearly confirm that both the main line and satellite emission can be
fit with two velocity components.

\begin{figure}
\centering
\resizebox{8.5cm}{!}{\includegraphics[angle=0]{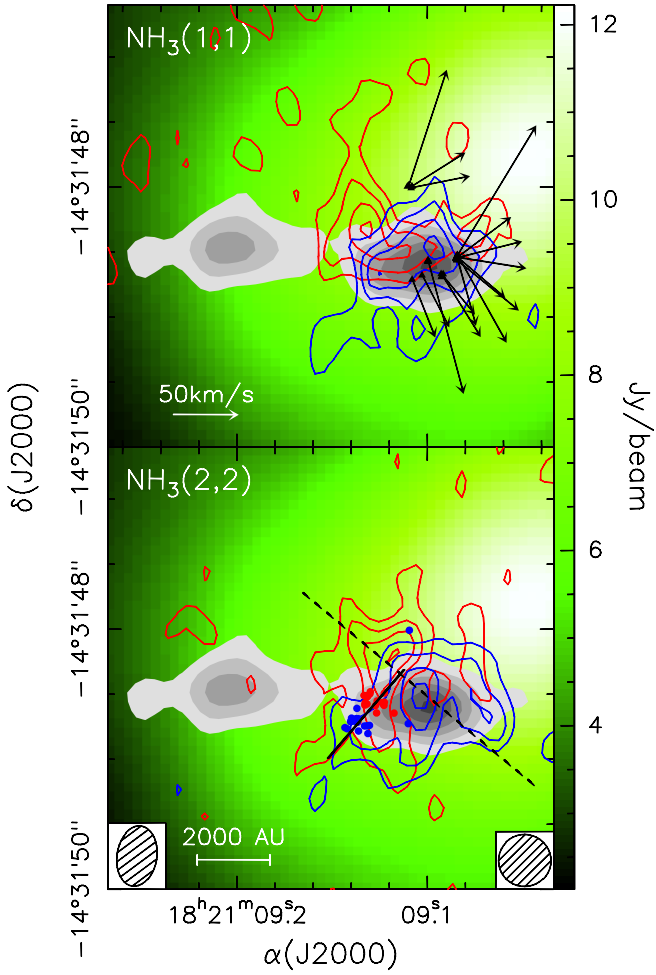}}
\caption{
 Maps of the two velocity components (see Fig.~\ref{fspamm}) identified in
 the \amm(1,1) (top) and (2,2) (bottom) lines. Blue and red contours are the
 maps obtained by averaging the main line emission, respectively, from 55.8
 to 60.6~\kms and from 60.6 to 63.0~\kms. The gray scale is the same map of
 the 6~cm continuum emission as in Fig.~\ref{fjsi}a, while the backgroud
 image (green tones) is the map of the 3.3~mm continuum emission obtained by
 F2008. The red and blue dots give the position of the red- and blue-shifted,
 respectively, methanol masers from S2010.
 The black triangles and arrows are the water maser spots and corresponding
 proper motions from S2010. The amplitude scale for the proper motions is given
 on the bottom left of the upper panel. The continuous and dashed line in the 
 lower panel indicate the axes of projection to produce
 the position-velocity plots of Figs.~\ref{fpvdisk}~and~\ref{fpvout}, respectively.
  The ellipses in the bottom left and
 right of the lower panel indicate, respectively, the HPBW of the 6~cm continuum and ammonia line
 maps.
}
\label{fvelcom}
\end{figure}

It is important to investigate the spatial distribution of these two
components, as done in Fig.~\ref{fvelcom}, where we compare the main line
\amm(1,1) and (2,2) emission integrated under the red and blue components,
with the radio jet and the core traced, respectively, by the centimeter and millimeter continuum emission.
For the sake of comparison, also the \meth\ and \wat\ masers are shown. We
have used different colors for the blue- and the red-shifted
\meth\ spots, to outline the rotation of the disk about a NE--SW axis.
 From this figure one sees that ammonia is tracing a compact bipolar
structure
roughly perpendicular to the \meth\ disk and slightly offset (by
$\sim$0\farcs4) to the NW with respect to it. Such a small offset
(less than the \amm\ map angular resolution)
might be
justified
by the fact that the gas density is higher to the NW of the disk than to
the SE.

The orientation and velocity of the ammonia structure being similar to the
(larger scale) NE--SW bipolar outflow detected by B2006 and \citet{Lop09}
suggests that both could be manifestations of the same ejection phenomenon.
Indeed, this possibility cannot be excluded, as ammonia has been observed in
association with bipolar jets in other massive sources \citep[see, e.g.,][]{Zha99}.
In this scenario, the ammonia gas would be preferentially entrained by
the flow on the NW side -- i.e., that facing the core -- where the
density is higher. This would explain why the \amm\ emission is slightly
skewed to the NW.

\subsubsection{The outflow scenario}
\label{soutf}

To verify the outflow hypothesis, one may estimate the relevant parameters from the
ammonia emission and compare them to those obtained for the larger-scale NE--SW
bipolar outflow.
We have computed the mass loss rate, momentum rate, and mechanical
luminosity by integrating the blue- and red-shifted emission in the \amm(1,1)
main line over the respective lobes, and dividing by the dynamical time
scale. In this calculation we have assumed that the line is optically thin for
reasons that will be explained in Sect.~\ref{sdisk}.
The dynamical time scale, obtained from the ratio between the length of the lobes, 
and the maximum speed in the line wings, is $\sim$1.6$\times$10$^3$~yr,
which is only a lower limit of the MYSO age. We use 0\farcs6 (or 0.018 pc) as the length of the lobes, estimated from the separation between the peaks of the blue- and red-shifted maps of the NH$_3$ emission (see Fig.\ref{fvelcom}).
For the maximum speed in the line wings, we take 6.3~\kms, which was evaluated taking the difference in velocity between the most blue- and red-shifted channels where ammonia emission is detected.
We caution that the derived dynamical time scale is not corrected for the (unknown) inclination, $i_{\rm o}$, of the outflow 
with respect to the line of sight. Such a correction, corresponding to the factor \ 
$\coth(i_{\rm o})$, becomes important if the outflow is close to the plane of the sky or 
the line of sight. 
The dynamical time scale measured for the NE--SW outflow  by \citet{Lop09} using CO single-dish observations
is \ $\approx$10$^4$~yr, significantly larger than the value determined from the \amm\ data and likely
close to the MYSO age.
For our estimates we used a gas temperature of 140~K, equal to
the maximum brightness temperature measured in the \amm\ line. This is
consistent with the rotational temperature (130$^{+36}_{-23}$~K) computed by
F2008 from the \mcn(5--4) lines. Note that reducing the temperature to, for example,
50~K would decrease the outflow parameters only by a factor $\sim$2.

We obtain $\dot{M}_{\rm out}\simeq3\times10^{-7}/X~M_\odot$~yr$^{-1}$,
$\dot{P}_{\rm out}\simeq2\times10^{-11}/X~M_\odot$~\kms~yr$^{-1}$,
$\dot{E}_{\rm out}\simeq7\times10^{-7}/X~L_\odot$, where $X$ is the relative
abundance of \amm\ with respect to H$_2$. For a fiducial $X=4\times10^{-8}$
(see e.g., Van~Dishoek et al.\ 1993, S\'anchez-Monge et al.\ 2013c and references therein),
the value of $\dot{P}$ ($5\times10^{-4}$~M$_\odot$~\kms~yr$^{-1}$)
 is only a few times less than that estimated for the
NE--SW outflow by \citet{Lop09} from the $^{13}$CO(2--1) line
($0.82$--$2\times10^{-3}$~M$_\odot$~\kms~yr$^{-1}$).  Such a discrepancy could
be explained by the uncertainty on the ammonia abundance and we thus conclude
that ammonia might be tracing the root of the larger scale CO outflow
oriented NE--SW.

\subsubsection{The two-clump scenario}
\label{stwoclu}

\begin{figure}
\centering
\includegraphics[width=8.5cm]{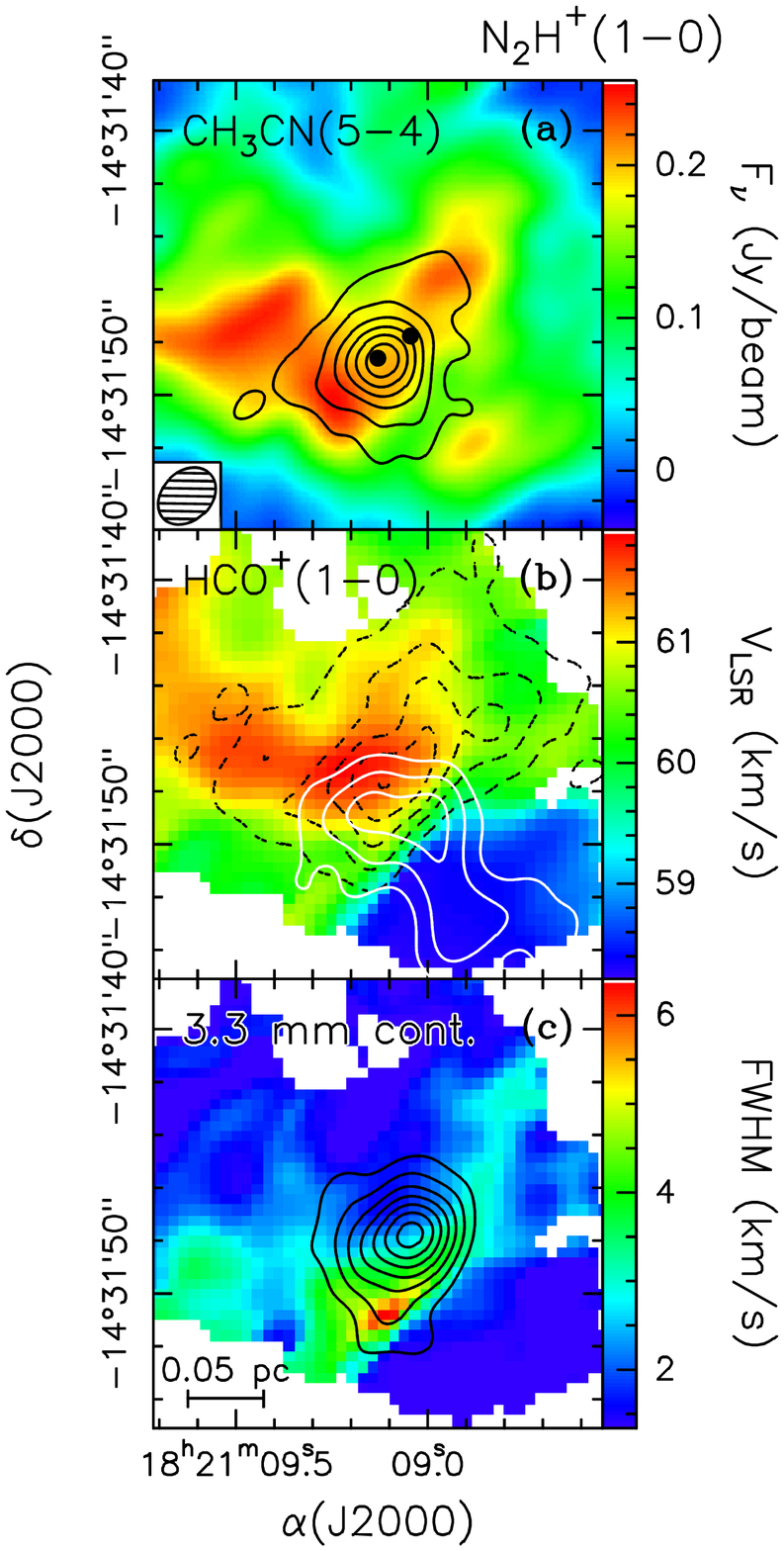}
\caption{
 Obtained from the data of F2008:
 {\bf a.} Map of the \mcn(5--4) line emission (contours) tracing the HMC,
 overlaid on a map of
 the \nthp(1--0) emission (from F2008) averaged over the main line (color
 scale). The black dots indicate the positions of the centimeter sources ``a''
 (alias mm-core) and
 ``b'' (alias HMC), detected by \citet{Zap06}.
 \ {\bf b.} Maps of the blue- (solid white contours) and red-shifted (dashed
 black) emission of the \hcop(1--0) line overlaid on a map of the \nthp(1--0)
 line velocity. The \hcop(1--0) maps are derived using the most extreme velocity 
 channels (1.68~\kms\ wide) at which  
 emission was detected by F2008, centered at \ 55.8~\kms\ and 62.5~\kms\ for the
 blue- and red-shifted line, respectively.
\ {\bf c.} Map of the 3.3~mm continuum emission (contours) tracing the mm-core,
 overlaid on a map of the \nthp(1--0) line FWHM.
}
\label{fnthp}
\end{figure}

We also consider another explanation for the two ammonia components, namely
that they are tracing two distinct, nearby clumps with slightly different
velocities. To investigate this possibility, we have analyzed the \nthp(1--0)
data of F2008 again. \nthp\ is found in close association with \amm\ and the OVRO
maps obtained by F2008 are sensitive to more extended emission than the
higher-resolution ammonia maps obtained by us. We have fit the \nthp\ spectra
taking into account the hyperfine splitting of the transition and created
maps of the average emission over the main line, peak velocity, and line FWHM.
These are shown, respectively, in Figs.~\ref{fnthp}a, \ref{fnthp}b, and
\ref{fnthp}c. Overlaid on these, we also show contour maps of the \mcn(5--4)
line emission, \hcop(1--0) blue- and red-shifted emission, and 3.3~mm
continuum emission.

The most striking feature is the sharp change in velocity seen in
Fig.~\ref{fnthp}b along a NE--SW line, which looks consistent with the
bipolar structure seen in \hcop. This result weakens the outflow
interpretation significantly, because \nthp\ is a well known tracer of dense
gas, but is only exceptionally detected in outflows 
\citep[see the case of L1157;][]{tob11}. Moreover, for a bipolar outflow one would
not expect such a sharp change in velocity because the line peak should shift
from red to blue velocities gradually, going from one lobe to the other. We
thus believe that the existence of two clumps is at least as likely as the
outflow scenario. Indeed, looking at Fig.~\ref{fnthp}a, one sees that while
most of the emission outlines a big, complex clump to the NE (where the
compact \mcn\ core is located), a minor peak of emission is also seen to the
SW, possibly tracing another less massive clump.

Let us hence assume for the moment that we are dealing with two nearby clumps
moving with different velocities along the line of sight. The question is
whether these two clumps are physically interacting or only overlapping in
projection on the plane of the sky. Comparison between Figs.~\ref{fnthp}c and
\ref{fnthp}b reveals that the line FWHM attains a maximum just along the line
where the sudden velocity change is seen. The fact that such a maximum
coincides with a tail of millimeter continuum emission (contours in
Fig.~\ref{fnthp}c) seems to confirm that the FWHM increase is not due to
two velocity components overlapping along the line of sight, but to a real
enhancement of the column density. Although hazardous at this stage, one might
even speculate that the formation of the mm-core has been triggered by the
collision of the two clumps, as it has been observed in other regions 
\citep{Dua11,Hen13}.

Clearly, further observational evidence is needed before drawing any conclusion
on the nature of the velocity field in this region. However, if the two-clump
hypothesis were correct, this could imply that the NE--SW bipolar outflow
claimed by B2006 and \citet{Lop09} does not exist.

The scenarios proposed here and in the Sect.~\ref{soutf} will be further
discussed in Sect.~\ref{sscen}.

Before that, however, we wish to analyze the velocity field of the
ammonia gas in some better detail, searching for a connection with the
\meth\ maser disk proposed by S2010.

\subsection{Ammonia from the \meth\ maser disk}
\label{sdisk}

Our data make it possible to find out whether the maser disk proposed by
S2010 is detected also in the ammonia emission. The most direct approach
consists in studying the gas velocity along the disk plane by means of a
position--velocity plot of the \amm\ emission. Since the \amm(1,1) main line
is slightly blended with the inner satellites (see Fig.~\ref{fspamm}) and the
(1,1) transition traces more extended (and colder) regions than the (2,2)
line, we have chosen the latter for our purposes. The plane of the disk is
assumed to lie along the direction joining the centers of the red- and
blue-shifted maser clusters (see e.g., Fig.~\ref{fvelcom}), i.e., along a P.A.
of 140\degr.

\begin{figure}
\centering
\resizebox{8.5cm}{!}{\includegraphics[angle=-90]{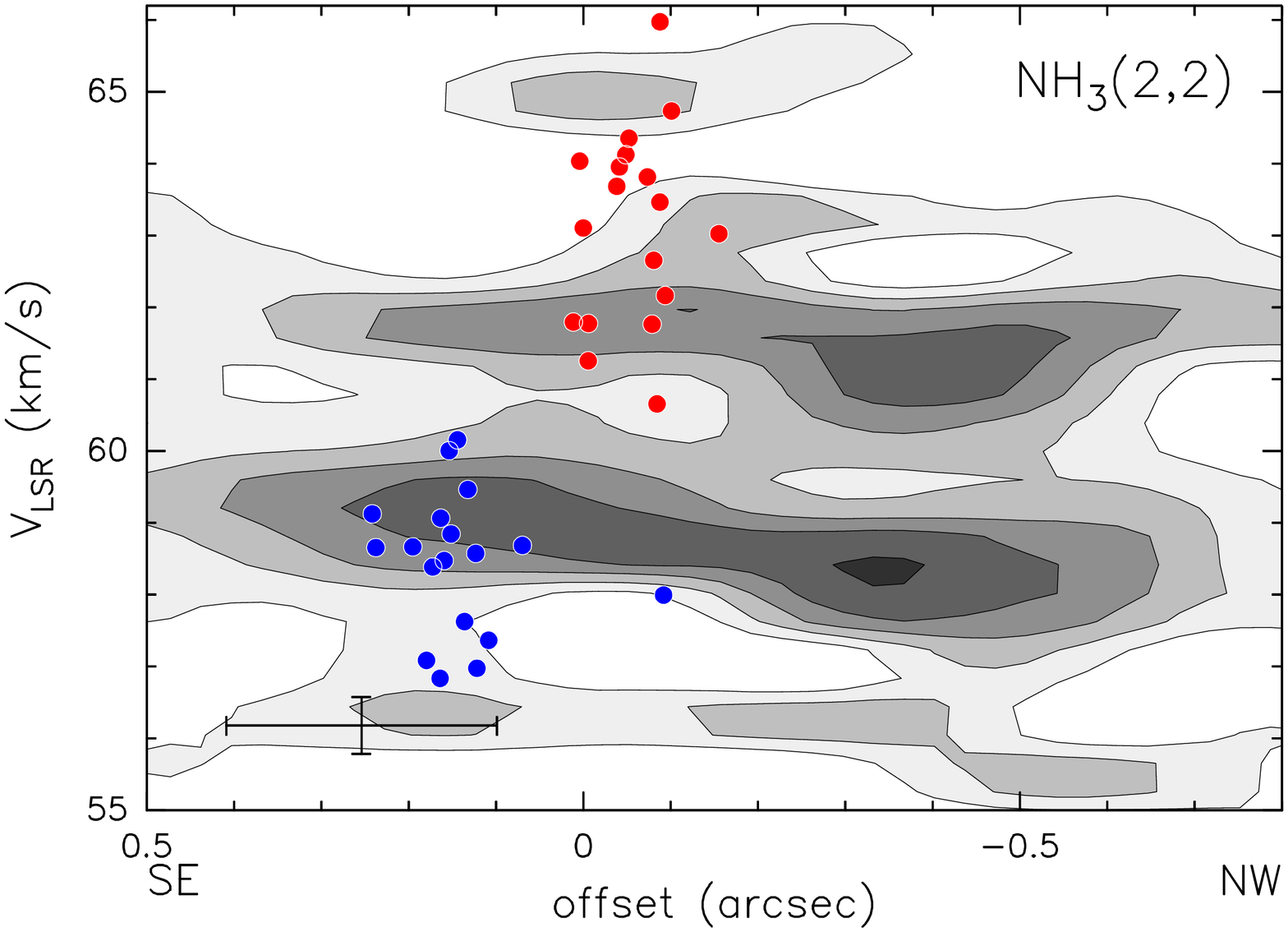}}
\caption{
Position--velocity plot of the \amm(2,2) line emission along a direction
crossing the centers of the red- and blue-shifted \meth\ maser clusters
(P.A.=140\degr). Contour levels range from 2 to 6~mJy/beam in steps of
1~mJy/beam. The red and blue circles represent the methanol maser spots.
The cross in the bottom left indicates the angular and spectral resolutions.
}
\label{fpvdisk}
\end{figure}

\begin{figure}
\centering
\resizebox{8.5cm}{!}{\includegraphics[angle=-90]{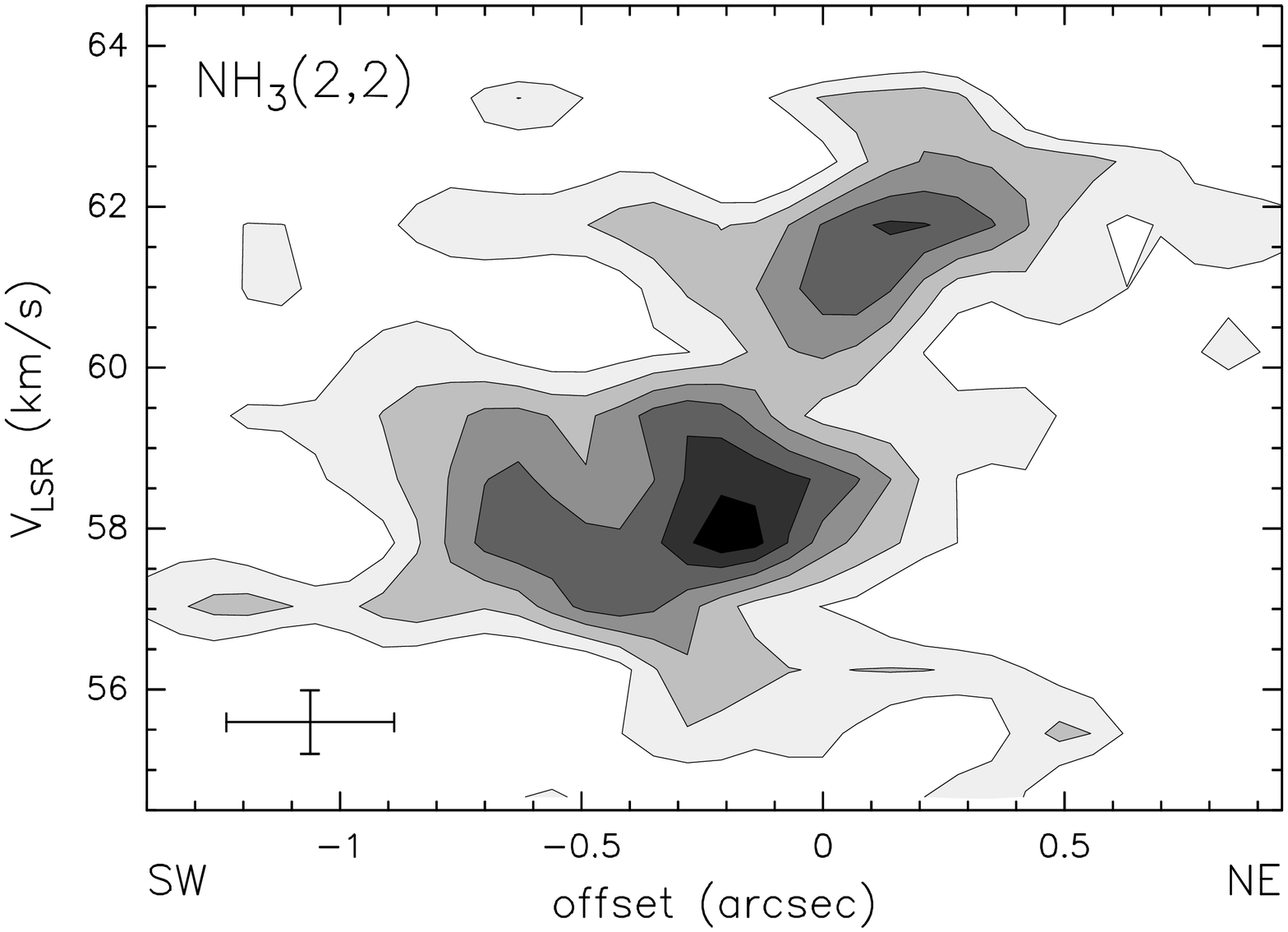}}
\caption{
Position--velocity plot of the \amm(2,2) line emission along a direction
perpendicular to that of Fig.~\ref{fpvdisk} (i.e., with P.A.=50\degr) and
passing at an offset of 0\farcs4 to the NW of the \meth\ masers.
Contour levels range from 2 to 7~mJy/beam in steps of
1~mJy/beam.
The cross in the bottom left indicates the angular and spectral resolutions.
}
\label{fpvout}
\end{figure}

The position-velocity plot of the \amm(2,2) main line is shown in
Fig.~\ref{fpvdisk}, where we also report the points representing the
\meth\ maser spots. One sees that ammonia emission is indeed detected
at the same positions and velocities as the two groups of masers, consistent
with emission from a rotating disk/ring. However, the emitting region is not
limited to the locations of the masers, but extends up to an offset of
$\sim$0\farcs4 to the NW. Here the \amm\ emission appears to attain its
maximum, both at blue- and red-shifted velocities. Such an offset
between the methanol masers center and the location of the two ammonia
velocity components was already pointed out in Sect.~\ref{sammo} (see
also Fig.~\ref{fvelcom}). In fact, the position--velocity plot along the
direction perpendicular to the disk plane and passing through the
ammonia peak (i.e., at an offset of 0\farcs4 to the NW of the maser center)
clearly outlines the two velocity components, as shown in Fig.~\ref{fpvout}.

We conclude that the methanol maser disk is detected also in the ammonia
emission, although quite faint.

\section{A model for the \Gb\ region}
\label{sscen}

Before
proposing an interpretation for
the \Gb\ star-forming
region, it is worth summarizing the observational findings obtained in this and
previous studies:
\begin{itemize}
\item A compact ($\sim$0.04~pc) dusty core is detected at millimeter
 wavelengths. Hot-core tracers such as \mcn\ and \amm\ peak at a
 position offset by $\sim$1\farcs3 ($\sim$0.022~pc) to the SE
 from the core center.
\item Three mid-IR sources are seen within $\sim$15\arcsec\ from the HMC, but
 only that associated with the HMC appears to dominate the bolometric luminosity
 of the whole region ($\sim$10$^4~L_\odot$, corresponding to a $\sim$13~$M_\odot$
 ZAMS star).
\item 
 The detection of two outflows extending over $\sim$20\arcsec\ (0.35~pc)
 has been claimed by various authors, one directed NE--SW (B2006;
 L\'opez-Sepulcre et al. 2009) and the other SE--NW (B2006; F2008).
\item
 Our ammonia observations have detected two velocity components arising from
 a compact region ($\sim$0\farcs6 or 0.01~pc). We propose that these might
 be tracing either the root of the bipolar outflow directed NE--SW, or the
 region of overlap between two clumps moving with different velocities
 along the line of sight.
\item A radio jet directed E--W is imaged from 6 to 1.3~cm. While the 6~cm
 continuum is dominated by non-thermal (possibly synchrotron) emission,
 at 1.3~cm compact free-free emission appears to arise very close to the
 center of the group of \meth\ maser spots detected by S2010.
 A 10~$M_\odot$ ZAMS star is sufficient to explain the observed flux in
 terms of ionization by the stellar Lyman continuum -- although ionization
 by shocks along the radio jet cannot be excluded.
\item The \wat\ maser proper motions measured by S2010 seem
 to trace expansion to the west, along the western border of the radio jet.  The
 \meth\ maser spots cluster in two groups that appear to rotate about the
 free-free continuum peak, with a rotation axis oriented NE--SW and an
 equilibrium mass of $\sim$12~$M_\odot$ (from S2010, after
 correcting for the new distance estimate). A similar velocity field is
 seen also in the ammonia lines.
\end{itemize}

\begin{figure}
\centering
\includegraphics[width=9cm]{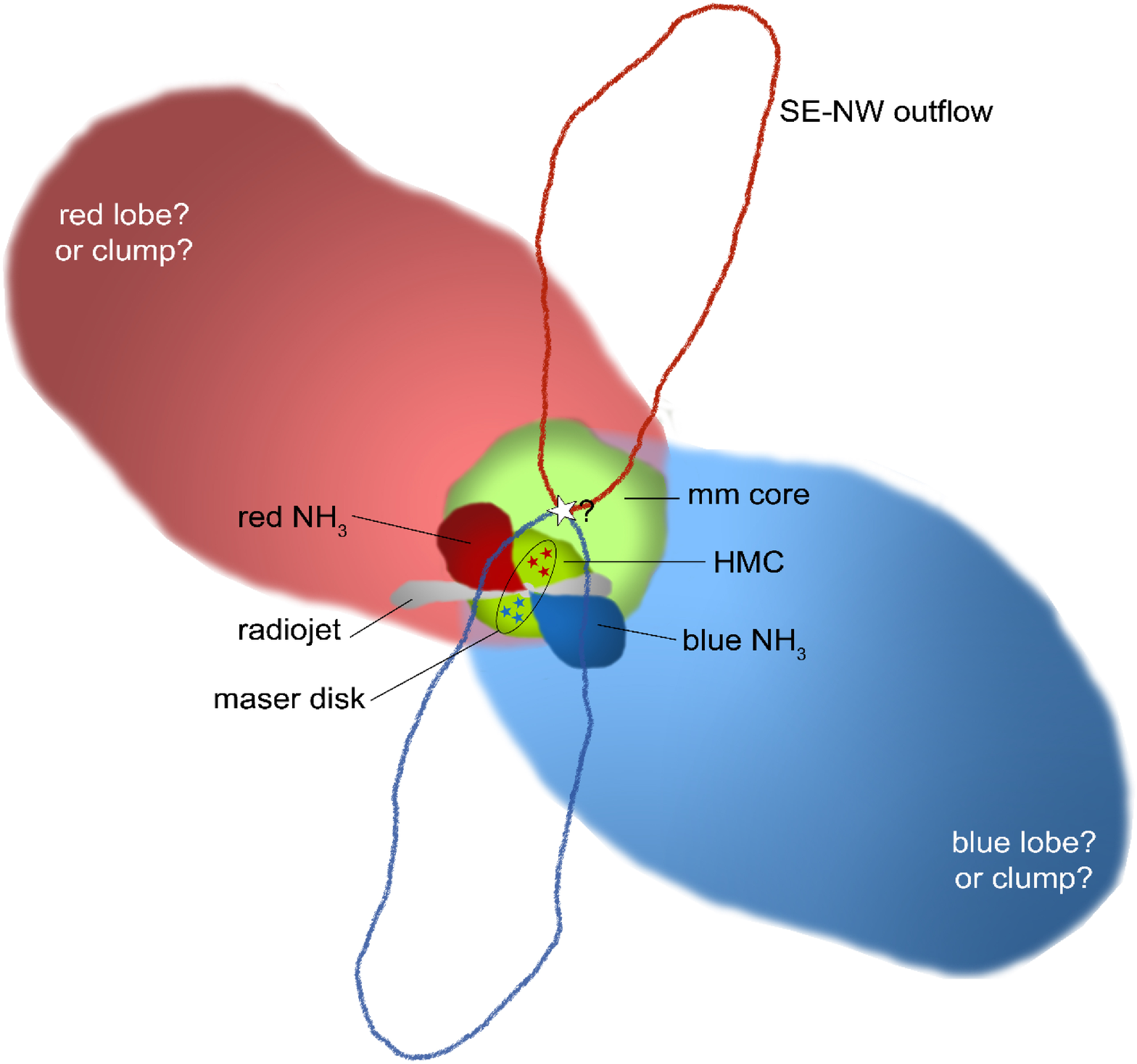}
\caption{
Sketch of the scenarios proposed to explain all the features observed in
the \Gb\ region.
Note that the relative sizes of the different structures (disk, jet, outflow, etc.)
have been arbitrarily chosen to make the figure more readable.
}
\label{fsketch}
\end{figure}

We have proposed two scenarios to interpret the \amm\ and \nthp\ observations
and in the following we will refer to them as ``case~A'', for that depicted
in Sect.~\ref{soutf} (ammonia emission associated with the NE--SW bipolar
outflow) and ``case~B'', for that illustrated in Sect.~\ref{stwoclu} (ammonia
emission from two overlapping clumps with different velocities).
Figure~\ref{fsketch} shows a sketch that summarizes both scenarios.

Neither case~A nor case~B can be explained with only one dominant MYSO in the
\Gb\ region. In fact, case~B implies the existence of a radio jet associated
with a \meth\ maser disk (see Sect.~\ref{sjet}), plus the SE--NW outflow
detected by B2006 and F2008. On the one hand, methanol masers have never been
found in association with low-mass YSOs; on the other, the momentum rate of
the SE--NW outflow is typical of MYSOs. We thus conclude that both the radio
jet and the outflow must be powered by MYSOs. In this scenario the disk-jet
system is associated with the HMC, while the outflow might be originating
from a young massive protostar deeply embedded in the mm-core.

The situation of case~A is more complex. Here we have the disk-jet system,
the SE--NW outflow, and the NE--SW outflow that we detect in ammonia on a
small scale.  Like in case~B, the (so far undetected) MYSO powering the
SE--NW outflow might be lying inside the mm-core. As for the radio jet and
NE--SW outflow, we propose that both could be associated with the same MYSO.
The problem with this hypothesis are the different directions of the E--W jet
and NE--SW outflow. This discrepancy  can be explained
if the jet/outflow is undergoing precession,
because the direction of the jet/outflow axis would change from the small to
the large scale, as in the case of the massive protostar IRAS\,20126+4104
\citep{She00,Ces05}.  Alternatively, the apparent direction of the radio jet
might be affected by the density distribution around the YSO powering the
jet, as the density appears to be much higher to the west where the core is
located. As already noted in Sect.~\ref{sammo}, such an asymmetric
distribution could also justify the small offset observed between the
\meth\ masers and the geometrical center of the \amm\ flow (see
Fig.~\ref{fvelcom}).

Is the case-A scenario compatible with the physical parameters estimated by us?
We note that the values of the stellar mass (10--13~$M_\odot$) derived from the
bolometric luminosity, \meth\ maser rotation velocity, and free-free 1.3~cm
continuum, are all in good agreement with each other. 
Using models of outflow evolution recently presented by \citet{Dua13},
a star of final mass of \ 10--15~$M_\odot$, evolved to the stage of a HMC,
should emit outflows with typical momentum rates of a few 10$^{-4}~M_\odot$~\kms~yr$^{-1}$.
Considering the uncertainty on the inclination angle of the outflow, such a value agrees well
with the momentum rate, $5\times10^{-4}$~M$_\odot$~\kms~yr$^{-1}$, we have derived 
in Sect.~\ref{soutf} for the \amm\ flow. 
%

Our conclusion is that the observational evidence collected so far is
insufficient to decide between case~A and B, although we tend to favor the
latter because case~A implies the existence of a NE--SW outflow traced also
by the \nthp(1--0) line: to our knowledge, this would be the second case ever
of \nthp\ tracing an outflow -- and the very first case for outflows from MYSOs.

\section{Summary and conclusions}
\label{conclu}

We performed ammonia line and radio and IR continuum observations of the
\Gb\ star-forming region with subarcsecond resolution, to shed light on the
structure and velocity field of the
mm-core and
HMC nearby the \meth\ maser disk imaged
by S2010, and establish the origin of the two bipolar outflows observed in the
region. We discover a radio jet oriented E--W and centered on the
\meth\ masers, whose western lobe appears to expand to the west, impinging
against the HMC, as suggested by the \wat\ maser proper motions measured by
S2010. The radio continuum from the jet has negative spectral index between 6
and 2~cm, indicating non-thermal continuum (synchrotron) emission, with the
sole exception of a compact free-free continuum source close to the \meth\
masers.

Our 24.5~$\mu$m image complemented by Herschel continuum data from the
Hi-GAL survey indicate that the bolometric luminosity of the region is
dominated by the emission from the neighborhoods of the HMC
and mm-core.
The stellar mass implied by the observed luminosity ($\sim$13~$M_\odot$) is
in good agreement with both the equilibrium mass of the \meth\ maser disk
($\sim$12~$M_\odot$), and the stellar mass ($\ga$10~$M_\odot$) needed to
explain the free-free continuum emission -- assuming the latter is due to
ionization by stellar photons.

We find that the ammonia emission presents a bipolar structure consistent
(on a smaller scale) in direction and velocity with that of the NE--SW
bipolar outflow reported by B2006 and \citet{Lop09}. After analyzing the
\nthp(1--0) observations of F2008 again, we conclude that two scenarios are possible.
Both imply the existence of two MYSOs, but differ in the
interpretation of the NE--SW bipolar structure. In one case this is a bipolar
outflow (in agreement with B2006 and \citet{Lop09}), which represents the
prosecution on the large scale of the small-scale radio jet.
In the other
case, the bipolarity is interpreted as two overlapping clumps moving with
different velocities along the line of sight. 
We slightly favor the second hypothesis, because the first would
imply that the NE--SW outflow is detected also in the \nthp(1--0) line emission,
which appears unlikely. Discriminating between the two scenarios requires
observations with higher sensitivity than our VLA ammonia images and we
believe that the Atacama Large Millimeter Array will be the right instrument
to establish the nature of the NE--SW bipolar structure and investigate the
circumstellar disk associated with the methanol masers.

\begin{acknowledgements}
We thank the anonymous referee for his constructive critcisms and
the suggestion to analize the \nthp\ data by F2008 again, which helped us
to shed new light on this complex region.
R.S.F. acknowledges T. Usuda, T. Inagaki , S. S. Hayashi, and H. Shinnaga 
for their help with the Subaru observations and data reduction.
We are greatful to Beatriz S\'anchez Monge and 
Rafael Delgado Romero for preparing the sketch shown in Fig.~\ref{fsketch}.
A.~Sanna acknowledges the financial support by the European Research Council 
for the ERC Advanced Grant GLOSTAR under contract no. 247078.
\end{acknowledgements}

\bibliographystyle{aa} 
\bibliography{biblio} 

\end{document}